\documentstyle[aps,preprint,prl,epsfig]{revtex}

\draft
\newcommand{\ba}{\begin{eqnarray}}
\newcommand{\ea}{\end{eqnarray}}
\newcommand{\bmath}{\begin{mathletters}}
\newcommand{\emath}{\end{mathletters}}
\newcommand{\ban}{\begin{eqnarray*}}
\newcommand{\ean}{\end{eqnarray*}}
\newcommand{\tl}{\tilde{\ell}}
\begin{document}

\title{Test of Nuclear Wave Functions for Pseudospin Symmetry}
\author{J.N. Ginocchio$^{1}$ and A. Leviatan$^{2}$}
\address{$^{1}$~Theoretical Division, Los Alamos National Laboratory, Los
Alamos, New Mexico 87545, USA}
\address{$^{2}$~Racah Institute of Physics, The Hebrew University,
Jerusalem 91904, Israel}
\date{\today}

\maketitle

\begin{abstract}
Using the fact that pseudospin is an approximate symmetry of the Dirac
Hamiltonian with realistic scalar and vector mean fields, we derive the
wave functions of the pseudospin partners of eigenstates of a realistic Dirac
Hamiltonian and compare these wave functions with the wave functions of the
Dirac eigenstates.
\end {abstract}


\newpage

Pseudospin doublets were introduced more than thirty years ago into nuclear
physics to accommodate an observed near degeneracy of certain normal-parity
shell-model orbitals with non-relativistic quantum numbers
($n_r$, $\ell$, $j = \ell + 1/2)$ and
($n_{r}-1, \ell + 2$, $j = \ell + 3/2$)
where $n_r$, $\ell$, and $j$ are the
single-nucleon radial, orbital, and total  angular momentum quantum
numbers, respectively \cite{aa,kth}. The doublet structure,
$j = \tilde{\ell}\pm \tilde s$, is expressed
in terms of a ``pseudo'' orbital angular momentum
$\tilde{\ell}$ = $\ell$ + 1 coupled to a
``pseudo'' spin, $\tilde s$ = 1/2.
This pseudospin ``symmetry''
has been used to explain features of deformed nuclei \cite{bohr},
including superdeformation \cite{dudek}
and identical bands \cite{twin,stephens}. 
Although the observed reduction in pseudo spin-orbit splitting follows
from nuclear relativistic mean-fields \cite{draayer},
only recently has the pseudospin ``symmetry'' been shown to arise from a
relativistic symmetry of the Dirac Hamiltonian \cite {gino,ami}.

The Dirac Hamiltonian, $H$,
with an external scalar, $V_S$, and vector, $V_V$, potentials is 
invariant under a SU(2) algebra for $V_S$ = $V_V $ + constant, 
leading to pseudospin symmetry in nuclei \cite{ami}. 
The pseudospin generators, $\hat{\tilde{S}}_{\mu}$,
which satisfy $[\,H\,,\, \hat{\tilde{S}}_{\mu}\,] = 0$ in the symmetry 
limit, are given by
\ba
{\hat{\tilde {S}}}_{\mu} =
\left (
\begin{array}{cc}
\hat {\tilde s}_{\mu} &  0 \\
0 & {\hat s}_{\mu}
\end{array}
\right )
= \left (
\begin{array}{cc}
U_p\, {\hat s}_{\mu} \, U_p & 0 \\
0 & {\hat s}_{\mu}
\end{array}
\right )
\label{psgen}
\ea
where
${\hat s}_{\mu} = \sigma_{\mu}/2$ are the usual spin generators,
$\sigma_{\mu}$ the Pauli matrices, and
$U_p = \, {\mbox{\boldmath $\sigma\cdot p$} \over p}$ is the
momentum-helicity unitary operator introduced in \cite {draayer}. If,
in addition, the potentials are spherically symmetric,
$V_{S,V}({\mbox{\boldmath $r$}})
= V_{S,V}({{ r}})$, the Dirac Hamiltonian has an additional invariant
SU(2) algebra, $[\,H\,,\, \hat{\tilde{L}}_{\mu}\,] = 0$, with the
pseudo-orbital angular
momentum operators given by
$\hat{\tilde{L}}_{\mu} =
\left ( {\hat {\tilde \ell_{\mu}} \atop 0 }
{ 0 \atop { {\hat \ell}_{\mu}} }\right )$. Here
$\hat {\tilde \ell}_{\mu} = U_p\, {\hat \ell}_{\mu}$ $U_p$,
${\hat\ell}_{\mu}= \mbox{ \boldmath $r$}\times \mbox{ \boldmath $p$}$,
while ${\hat j}_{\mu} = {\hat {\tilde \ell}}_{\mu}
+ {\hat {\tilde s}_{\mu}} = U_p\,(\, {\hat \ell}_{\mu}
+ {\hat s}_{\mu} \, )\, U_p = {\hat \ell}_{\mu} + {\hat s}_{\mu} $.
The eigenfunctions of the Dirac Hamiltonian
are also eigenfunctions of the Casimir operator of this algebra,
$\mbox{\boldmath $\hat{\tilde{L}}\cdot \hat{\tilde{L}}$}\,
\vert\tilde {n}_r,\, {\tilde \ell},\, j,\, m \,\rangle
= {\tilde\ell} ({\tilde \ell} + 1) \vert\tilde {n}_r,\,
{\tilde \ell},\, j,\, m \,\rangle $,
where we have used a coupled basis,
${\mbox{\boldmath $j$}} = {\mbox{\boldmath $\tilde\ell$}} +
{\mbox{\boldmath $\tilde{s}$}}$, and set $\hbar = c =1$.
Here $j$ is the eigenvalue of the total
angular momentum operator ${\hat J}_{\mu} = {\hat {\tilde L}}_{\mu}
+ {\hat{\tilde S}}_{\mu},\
\mbox{\boldmath ${\hat J}\cdot{\hat J}$}\,
\vert\tilde {n}_r,\, {\tilde \ell},\, j,\, m \,\rangle
= j(j + 1)\vert\tilde {n}_r,\, {\tilde\ell},\, j,\, m \,\rangle $, $m$
is the eigenvalue of ${\hat J}_z$ and $\tilde {n}_r$ is the pseudoradial
quantum number which we define below.

In the pseudospin symmetry limit,
the eigenstates of the Dirac  Hamiltonian
in the doublet $j = {\tilde {\ell}} \pm 1/2$ are degenerate,
and are connected by the pseudospin generators
$\hat{\tilde{S}}_{\mu}$:
\ba
\hat{ \tilde{S}}_{\mu}\,
\vert\, \tilde {n}_r,\  { \tilde {\ell}},\  j_i ,\ m_i\,\rangle =
\sum_{j_f,\, m_f} A_{j_f,m_f, j_i,m_i}\, \vert\,\tilde {n}_r,\
{ \tilde{\ell}},\ j_f,\  m_f\,\rangle ~.
\label {pspeig}
\ea
Here $A_{j_f,m_f, j_i,m_i} = (-1)^{{1\over2}-m_f+\tilde
{\ell}} \sqrt{3(2j_i+1)(2j_f+1)\over 2}
\left ( {j_f \atop -m_f } {1\atop \mu} { j_i \atop m_i}\right ) \
\left
\{ {{1\over 2} \atop j_f } { \tilde {\ell} \atop 1} { j_i \atop
{1\over 2}}\right \}$, where the symbols are
Wigner 3-j and 6-j symbols respectively.
However, in the exact pseudospin limit, $V_S=-V_V$, there are no bound
Dirac valence states. For nuclei to exist the pseudospin symmetry must
therefore be broken. Nevertheless, realistic mean fields involve
an attractive scalar potential and a repulsive vector potential of
nearly equal magnitudes, $V_S \sim - V_V$, and calculations
in a variety of nuclei confirm the existence of an approximate
pseudospin symmetry in the energy spectra \cite{gino2,ring,arima}.
Since pseudospin symmetry is broken, the
pseudospin partner produced by the raising and lowering
operators acting on an eigenstate will not necessarily be an eigenstate.
The question is how different is the pseudospin
partner from the eigenstate with the same quantum numbers? As noted,
energy splittings suggest that the breaking of pseudospin symmetry is
small, but is the breaking in the eigenfunctions small as well? 
While previous studies have compared the lower components
of the Dirac wave functions of the two states in the doublet
\cite{gino2,ring,arima},
it is the behaviour of the upper components which is of most interest
since they dominate the Dirac eigenstates. The relativistic pseudospin 
symmetry has unique and interesting  features in the following sense.
First, the pseudospin generators  of Eq.~(\ref{psgen}) intertwine space
and spin, and thus lead  to an uncommon symmetry structure of doublets
with different radial wave functions. Second, since bound Dirac valence
states do not exist in the  symmetry limit, the
pseudospin properties of  realistic wave functions can not be determined
by perturbation theory. These aspects motivate the present study. 

To determine the pseudospin partners
we need to expand the Dirac eigenfunction into a spherical basis,
\bmath
\ba
\langle \,
\mbox{\boldmath $r$} \
\vert \, \tilde {n}_r,\ { \tilde {\ell}},
\ j= { \tilde {\ell}} + {1 \over 2}  ,\ m \, \rangle &=&
\left ( g_{\tilde {n}_r-1,\,{ \tilde {\ell}},\, j }(r)
[Y^{({{\tilde {\ell}}+1})}(\hat{r})\chi]^{(j)}_m \, , \,
if_{\tilde {n}_r,\,{ \tilde {\ell}},\, j }(r)
[Y^{({{\tilde{\ell}}})}(\hat{r})\chi]^{(j)}_m\right) \\
\langle \,
\mbox{\boldmath $r$} \
\vert \, \tilde {n}_r,\ { \tilde {\ell}},
\ j= { \tilde {\ell}} - {1 \over 2}  ,\ m \, \rangle &=&
\left ( g_{\tilde {n}_r,\,{ \tilde {\ell}},\, j }(r)
[Y^{({{\tilde{\ell}}-1})}(\hat{r})\chi]^{(j)}_m \, , \,
if_{\tilde {n}_r,\,{ \tilde {\ell}},\, j }(r)
[Y^{({{\tilde{\ell}}})}(\hat{r}) \chi]^{(j)}_m\right),
\label {exp}
\ea
\emath
where $Y^{({{\tilde {\ell}}})}_{m_{\tl}}(\hat{r})$
is the spherical harmonic and
$\chi$ is the spin function. From this expansion we see that the
pseudoradial quantum number, $\tilde{n}_r$,
is the radial quantum number of the lower component of the Dirac
eigenfunction
\cite{rose} as well as the radial quantum number of the upper component
of the eigenstate with $j={ \tilde {\ell}}-1/2$.
Because the pseudospin generators ${\hat{\tilde {S}}}_{\mu}$ depend on
the unit momentum vector $\hat {p}$, we transform the
eigenfunctions to momentum space in order to calculate the effect of the
pseudospin generators on the eigenfunctions:
\bmath
\ba
\langle \,
\mbox{\boldmath $p$} \
\vert\,\tilde {n}_r,\  { \tilde {\ell}},\
j= { \tilde {\ell}} + {1 \over 2}  ,\ m\, \rangle &=&
\left (\tilde{g}_{\tilde {n}_r-1,\,{ \tilde {\ell}},\,j }(p)
[Y^{({{\tilde{\ell}}+1})}(\hat{p})
\chi]^{(j)}_m,i\tilde{f}_{\tilde {n}_r,\,{ \tilde {\ell}},\, j }(p)
[Y^{({{\tilde{\ell}}})}(\hat{p})\chi]^{(j)}_m\right) \\
\langle \,
\mbox{\boldmath $p$} \
\vert\,\tilde {n}_r,\  { \tilde {\ell}},\  j= { \tilde {\ell}} -
{1 \over 2}  ,\ m\, \rangle &=&
\left (\tilde{g}_{\tilde {n}_r,\,{ \tilde {\ell}},\, j }(p)
[Y^{({{\tilde{\ell}}-1})}(\hat{p})
\chi]^{(j)}_m,i\tilde {f}_{\tilde {n}_r,\,{ \tilde {\ell}},\, j }(p)
[Y^{({{\tilde{\ell}}})}(\hat{p}) \chi]^{(j)}_m\right).
\label {expp}
\ea
\emath
The corresponding spherical Bessel transforms of the radial wave
functions are given by
\bmath
\ba
\tilde{g}_{\tilde {n}_r-1,\, { \tilde {\ell}},\, j }(p) &=&
(-i)^{{\tilde {\ell}}+1}{\sqrt {2\over\pi}}\int_0^{\infty}
j_{_{{\tilde {\ell}}+1}}(pr)\
g_{\tilde {n}_r-1,\,{ \tilde {\ell}},\, j }(r)\ r^2 dr \qquad
j = \tilde{\ell} + {1\over2}\\
\tilde{g}_{\tilde {n}_r,\, { \tilde {\ell}},\, j }(p) &=&
(-i)^{{\tilde {\ell}}-1}{\sqrt {2\over\pi}}\int_0^{\infty}
j_{_{{\tilde {\ell}}-1}}(pr)\
g_{\tilde {n}_r,\, { \tilde {\ell}},\, j }(r)\ r^2 dr \qquad\;\;\;
j = \tilde{\ell} - {1\over2}\\
\tilde{f}_{\tilde {n}_r,\, { \tilde {\ell}},\, j }(p) &=&
(-i)^{{\tilde {\ell}}}{\sqrt {2\over\pi}}\int_0^{\infty}
j_{_{{\tilde {\ell}}}}(pr)\
f_{\tilde {n}_r,\, { \tilde {\ell}},\, j }(r)\ r^2 dr \qquad\qquad\;\;
j = \tilde{\ell} \pm {1\over2}.
\label {mom}
\ea
\emath
We then are able to derive
\ba
\hat{\tilde{S}}_{\mu}
\vert\,\tilde {n}_r,\  { \tilde {\ell}},\  j_i ,\ m_i\,\rangle  =
A_{j_i,m_i, j_i,m_i}\; \vert\,\tilde {n}_r,\
{ \tilde{\ell}},\  j_i ,\ m_i\,\rangle
+ \ A_{j_f,m_f, j_i,m_i}\;\vert\,\tilde {n}_r,\
{ \tilde {\ell}},\  j_f ,\ m_f\,\rangle^{psp} ~.
\label {psp}
\ea
Here the superscript $psp$ on the second term denotes the pseudospin
partner with $j_f \neq j_i$.
Even with pseudospin breaking, the pseudospin generators
do not change ${\tilde {\ell}}$.
In addition, from  Eq.~(\ref{psp}) we see that the first term
with $j_f=j_i$ is exactly equal to
the original eigenstate, independent of the amount of pseudospin
breaking. This follows
from the orthogonality of the spherical Bessel functions, $ {2
\over\pi}\int_0^{\infty}\ p^2 dp\ j_{_{{\tilde {\ell}}}}(pr)
j_{_{{\tilde {\ell}}}}(px) = {\delta(r-x)\over r^2}$. For the partner
with $j_f \neq j_i$, the wave function in coordinate space reads
\bmath
\ba
\langle \,
\mbox{\boldmath $r$} \
\vert\,\tilde {n}_r,\  { \tilde {\ell}},\
j= { \tilde {\ell}} + {1 \over 2}  ,\ m \, \rangle^{psp} &=&
\left (g_{{\tilde n}^{\prime}_r-1,\, { \tilde {\ell}},\, j }^{psp}(r)
[Y^{({{\tilde{\ell}}+1})}(\hat{r})\chi]^{(j)}_m \, , \,
if_{\tilde {n}_r,\, { \tilde {\ell}},\, j }^{psp}(r)
[Y^{({{\tilde{\ell}}})}(\hat{r})\chi]^{(j)}_m\right) \\
\langle \,
\mbox{\boldmath $r$} \
\vert\,\tilde {n}_r,\  { \tilde {\ell}},\
j= { \tilde {\ell}} - {1 \over 2}  ,\ m\, \rangle^{psp} &=&
\left ( g_{{\tilde n}^{\prime}_r,\, { \tilde {\ell}},\, j }^{psp}(r)
[Y^{({{\tilde{\ell}}-1})}(\hat{r})\chi]^{(j)}_m\, , \,
if_{\tilde {n}_r,\, { \tilde {\ell}},\, j }^{psp}(r)
[Y^{({{\tilde{\ell}}})}(\hat{r}) \chi]^{(j)}_m\right)
\label {exppsp}
\ea
\emath
where in general ${\tilde n}^{\prime}_r$ can differ from $\tilde{n}_r$
since the states with $j_f \neq j_i$ in Eq.~(\ref{psp}) are not
Dirac eigenstates. 
The expressions for 
$g_{{\tilde n}^{\prime}_r,\, { \tilde {\ell}},\, j-1 }^{psp}(r)$ 
with $j = \tilde{\ell} + {1\over2}$ and 
$g_{{\tilde n}^{\prime}_r-1,\, { \tilde {\ell}},\, j+1 }^{psp}(r)$ 
with $j = \tilde{\ell} - {1\over2}$ involve a double integral 
$-{2 \over\pi}\int_0^{\infty}p^2 dp\int_0^{\infty} x^2 dx$ 
over $j_{_{{\tilde {\ell}}-1}}(pr)j_{_{{\tilde {\ell}}+1}}(px)
g_{{\tilde n}_r-1,{ \tilde {\ell}},j }(x)$ and 
$j_{_{{\tilde {\ell}}+1}}(pr)j_{_{{\tilde {\ell}}-1}}(px)
g_{\tilde {n}_r,\, { \tilde {\ell}},\, j }(x)$
respectively.
The $p$-integration 
can be carried out 
and altogether we find \cite{sign},
\bmath
\ba
g_{{\tilde n}^{\prime}_r,\, { \tilde {\ell}},\, j-1 }^{psp}(r) &=&
g_{\tilde {n}_r-1,\, { \tilde {\ell}},\, j }(r)
- (2\tl +1) r^{\tl-1}\int_{r}^{\infty} {dx\over x^{\tl}}
g_{\tilde {n}_r-1,\, { \tilde {\ell}},\, j }(x) \qquad
j = \tilde{\ell} + {1\over2}\\
g_{{\tilde n}^{\prime}_r-1,\, { \tilde {\ell}},\, j+1 }^{psp}(r) &=&
g_{\tilde {n}_r,\, { \tilde {\ell}},\, j}(r)
- {(2\tl +1)\over r^{\tl+2}}\int_{0}^{r} dx \ x^{\tl+1}
g_{\tilde {n}_r,\, { \tilde {\ell}},\, j}(x) \qquad\quad\;\;\;
j = \tilde{\ell} - {1\over2}\\
f_{\tilde {n}_r,\, { \tilde {\ell}},\, j\mp 1 }^{psp}(r) &=&
f_{\tilde {n}_r,\, { \tilde {\ell}},\, j }(r)\qquad\qquad
\;\;\;\,\qquad\qquad\qquad\qquad\qquad\qquad\;\,
j = \tilde {\ell} \pm {1\over2}
\label{gpsp}
\ea
\emath

In the pseudospin limit $(V_S + V_V = \;{\rm constant})$
\ba
\vert\,\tilde {n}_r,\  { \tilde {\ell}},\  j ,\ m\, \rangle^{psp} &=&
\vert\,\tilde {n}_r,\  { \tilde {\ell}},\  j ,\ m\,\rangle\, ~.
\label {ps}
\ea
Since pseudospin symmetry is slightly broken in nuclei,
the pseudospin partner can differ from the Dirac eigenstate and it is
of interest to examine the deviations from the condition of Eq.~(\ref{ps}).
Dirac bound states satisfy $g_{\tilde {n}_r,\,{ \tilde {\ell}},\,
j=\tl\pm 1/2}\sim r^{\tl\pm 1}$ for
small $r$, and fall off exponentially
$\sim \exp(-\sqrt{M^2-E^2}\;r)$,
for large $r$ \cite{rose}, where $M$ is the nucleon mass 
and $E$ is the Dirac eigenenergy. Consequently, as seen from
Eq.~(8a), for the Dirac eigenstate with $j=\tl+1/2$, its pseudospin partner
$g_{{\tilde n}^{\prime}_r,\,{ \tilde {\ell}},\, j-1 }^{psp}$ has the
expected behavior for small and large $r$.
On the other hand, as seen from
Eq.~(8b), for the Dirac eigenstate with $j=\tl-1/2$, its pseudospin partner
$g_{{\tilde n}^{\prime}_r-1,\,{ \tilde {\ell}},\, j+1 }^{psp}\sim r^{\tl-1}$
for small $r$, and falls off as a power law $r^{-(\tl +2)}$ for large $r$.
As such it has a behavior which is very different from that
of a Dirac bound state with $j=\tl+1/2$. This asymmetry in the behavior
of the pseudospin partners of $j=\tl+1/2$ or $j=\tl-1/2$ Dirac eigenstates,
is evident in the analysis of nuclear wave functions presented below.
These realistic wave functions were obtained in a relativistic point
coupling model, and we refer the reader to \cite{gino2} for details on
the parameterization of the potentials, and the data
that has been used to fix it.

We first examine Dirac eigenstates with $j=\tl+1/2$ and wave functions
as in Eq.~(3a). Their partners with $j^{\prime}=j-1$ are obtained from
Eqs.~(8a,c). As an example, we consider the realistic relativistic
mean field Dirac eigenstates $\ 0d_{3/2},\ 1d_{3/2}$ ($\tl=1,\; j=3/2$)
for $^{208}$Pb \cite {gino2}. In
Fig.~1 we compare the spatial wave functions of these pseudospin partners,
$[P(0d_{3/2})]s_{1/2},\; [P(1d_{3/2})]s_{1/2}$
with the eigenstates, $1s_{1/2},\; 2s_{1/2}$. 
(The symbol $P$ means the $s_{1/2}$ partner of the $0d_{3/2}$ or 
$1d_{3/2}$ eigenstates). 
The lower components agree very well, which was noted previously
\cite{gino2,ring,arima}, except for some disagreement on the surface. For the
upper components the agreement is not as good in the magnitude but the shapes
agree well, with the number of radial nodes being the same
[${\tilde n}^{\prime}_r = {\tilde n}_{r}$ in Eq.~(7a)]. 
The agreement improves as the radial quantum number increases, 
which is consistent with 
the observed decrease in the energy splitting between the doublets 
\cite{gino,gino2}. 
As another example in the same category ($j=\tl+1/2$), we compare in
Fig.~2 the $[P(0h_{9/2})]f_{7/2}$
partner of the $0h_{9/2}$ eigenstate ($\tl=4,\;j=9/2$) with the $1f_{7/2}$
eigenstate. The radial wave functions have the same number of radial quantum
numbers and, again, the lower components agree better.

Next we examine the other category of Dirac eigenstates with
$j=\tl-1/2$ and wave functions as in Eq.~(3b). Their partners with
$j^{\prime}=j+1$ are obtained from Eqs.~(8b,c).
As an example, we consider the realistic relativistic mean field
eigenstates $0s_{1/2},\; 1s_{1/2},\; 2s_{1/2}$ ($\tl=1,\;j=1/2$)
for $^{208}$Pb \cite{gino2}.
The $0s_{1/2}$ eigenstate will have a partner which we denote by
$[P(0s_{1/2})]d_{3/2}$, but there is no
$d_{3/2}$ eigenstate at approximately the same energy as
the $\ 0s_{1/2}$ eigenstate, so there is no eigenfunction to compare to.
On the other hand, the $1s_{1/2}$ and $2s_{1/2}$ eigenstates
are almost degenerate with the $0d_{3/2}$ and $1d_{3/2}$
eigenstates respectively. In
Fig.~3 we compare the spatial wave functions of the pseudospin partners
$[P(1s_{1/2})]d_{3/2},\; [P(2s_{1/2})]d_{3/2}$
with the respective $0d_{3/2},\; 1d_{3/2}$ eigenstates.
These partners agree well with the eigenfunctions in
the interior but not on the nuclear surface. In fact, the partners do
not have the same number of nodes and do not fall off
exponentially but inversely as the cubic power, $r^{-3}$, in agreement
with the $r^{-(\tl+2)}$ behavior reported in Eq.~(8b).

The Dirac eigenstates with $\tilde {n}_r= 0$ and
$j = { \tilde {\ell}}-  1/2$ are special, because no eigenstates exist
with the quantum numbers of their partners, $\tilde {n}_r$ = 0 and
$j = { \tilde {\ell}} + 1/2$.
An example is given in Fig.~3a,b for ${ \tilde {\ell}} =1,\; j=1/2$.
For heavy nuclei these states with large $j$ are the ``intruder'' states.
Before the SU(2) algebra of pseudospin was discovered, these
states were discarded from the pseudospin scheme.
However, that is clearly not a valid procedure if pseudospin symmetry is a
symmetry of the Dirac Hamiltonian. As another example, we show in
Fig.~4a,b the radial wavefunction of the $[P(0f_{7/2})]h_{9/2}$ partner
of the $0f_{7/2}$ intruder state ($\tl=4,\;j=7/2$).
There is no quasi-degenerate $h_{9/2}$ eigenstate
to compare to. The upper component has the $r^{-6}$ falloff alluded to above.
Although both components have zero radial quantum number, they do not compare
well with the $0h_{9/2}$ eigenstate shown in Fig. 4c,d. In Fig.~4c,d we
show also the radial wavefunction of the $[P(1f_{7/2})]h_{9/2}$ partner
of the $1f_{7/2}$ state ($\tl=4,\;j=7/2$)
and compare it to the $0h_{9/2}$ eigenstate.
The upper component has again the $r^{-6}$ falloff and therefore does
not compare well on the surface. Also the number of radial quantum numbers
differ. The lower components agree better.

In summary, we have shown that the radial
wave functions of the upper components of the
$j =  {\tilde {\ell}} - 1/2$ pseudospin partner of the eigenstate with
$j =  {\tilde {\ell}} + 1/2$ is similar in shape to the
$j =  {\tilde {\ell}} - 1/2$ eigenstate but there is a difference in
magnitude. However, the $\tilde {n}_r\neq 0$ radial wave functions of the
upper components of the $j =  {\tilde {\ell}} + 1/2$ pseudospin partner of
the eigenstate with $j =  {\tilde {\ell}} - 1/2$ is not similar in shape
to the $j =  {\tilde {\ell}} + 1/2$ eigenstate. In fact these
wave functions approach $r^{\tilde {\ell}-1}$ rather than
$r^{\tilde {\ell}+1}$ for $r$ small, do not have the same number of
radial nodes as the eigenstates, and do not fall off exponentially as do
the eigenstates, but rather fall off as
$r^{-({{\tilde {\ell}}+2})}$. Furthermore, the pseudospin partners of the
``intruder" eigenstates, $\tilde {n}_r = 0$, also fall off as as
$r^{-({{\tilde {\ell}}+2})}$. We have confirmed that the radial
wave functions of the lower components of the pseudospin partners of
eigenstates of the Dirac Hamiltonian for $j =  {\tilde {\ell}} \pm 1/2$
are very similar to the eigenstates with the same
quantum numbers except for some differences on the surface. 

This research was supported in
part by the United States Department of
Energy under contract W-7405-ENG-36 and in part by the
U.S.-Israel Binational Science Foundation.\\

\pagebreak

\begin{figure}
\begin{center}
\end{center}
\noindent Figure 1. a) The upper component [$g(r)$] and
b) the lower component [$f(r)]$ in (Fermi)$^{-3/2}$ of the
$[P(0d_{3/2})]s_{1/2}$ partner of the $0d_{3/2}$ eigenstate compared to
the
$1s_{1/2}$ eigenstate.
c)~The upper component and d) the lower component of the
$[P(1d_{3/2})]s_{1/2}$ partner of
the $1d_{3/2}$ eigenstate compared to the $2s_{1/2}$ 
eigenstate for $^{208}$Pb
\cite{gino2}.
\end{figure}

\begin{figure}
\begin{center}
\end{center}
\noindent Figure 2. a) The upper component [$g(r)$] and
b) the lower component [$f(r)]$ in (Fermi)$^{-3/2}$ of the
$[P(0h_{9/2})]f_{7/2}$ partner of the $0h_{9/2}$ eigenstate compared to
the
$1f_{7/2}$ eigenstate for $^{208}$Pb \cite{gino2}.
\end{figure}

\begin{figure}
\begin{center}
\end{center}
\noindent Figure 3. a) The upper component [$g(r)$] and
b) the lower component [$f(r)$] in (Fermi)$^{-3/2}$ of the
$[P(0s_{1/2})]d_{3/2}$ partner of the $0s_{1/2}$ eigenstate.
c) The upper component and d) the lower component of the
$[P(1s_{1/2})]d_{3/2}$ partner of the $1s_{1/2}$ eigenstate compared to
the $0d_{3/2}$ eigenstate.
e) The upper component and f) the lower component of the
$[P(2s_{1/2})]d_{3/2}$ partner of the $2s_{1/2}$ eigenstate compared to
the
$1d_{3/2}$ eigenstate for $^{208}$Pb \cite{gino2}.
\end{figure}

\begin{figure}
\begin{center}
\end{center}
\noindent Figure 4. a) The upper component [$g(r)$] and
b) the lower component [$f(r)$] in (Fermi)$^{-3/2}$ of the
$[P(0f_{7/2})]h_{9/2}$ partner of the $0f_{7/2}$ eigenstate.
c) The upper component and d) the lower component of the
$[P(1f_{7/2})]h_{9/2}$ partner of the $1f_{7/2}$ eigenstate compared to
the
$0h_{9/2}$ eigenstate for $^{208}$Pb \cite{gino2}.
\end{figure}


\begin{thebibliography}{99}
\bibitem{sign} 
We choose the overall phase of the pseudospin partner wavefunction 
such that it agrees with the eigenstate wavefunction 
with the same quantum numbers.
\end{thebibliography}
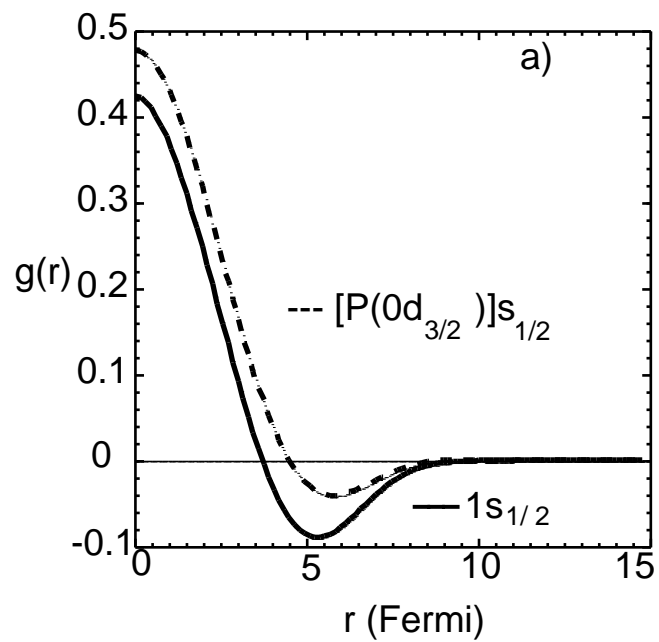
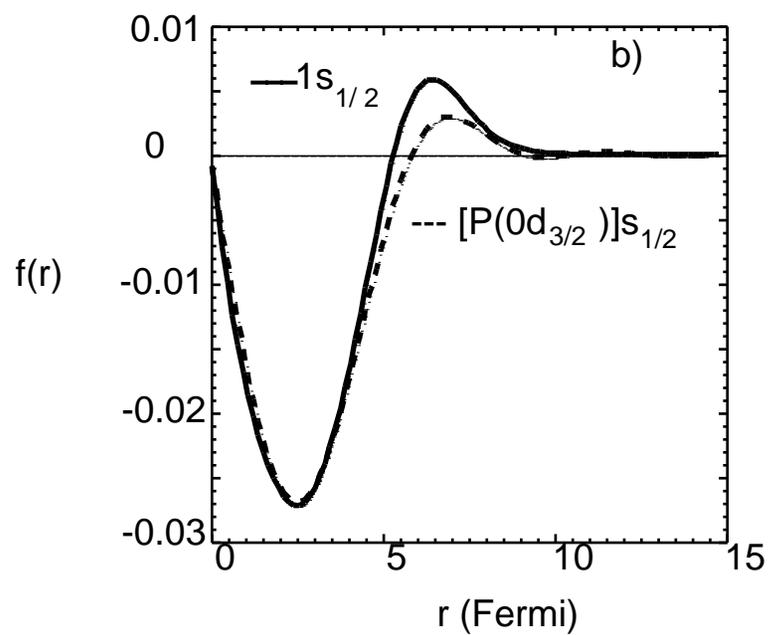
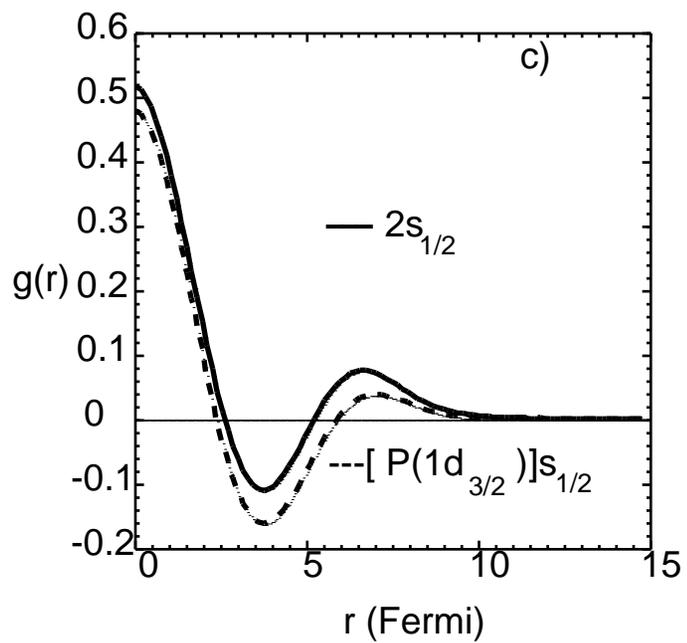
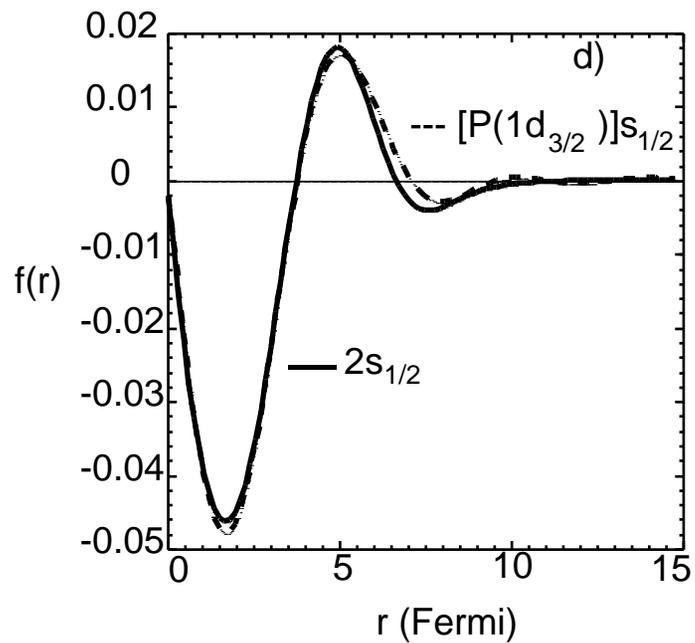
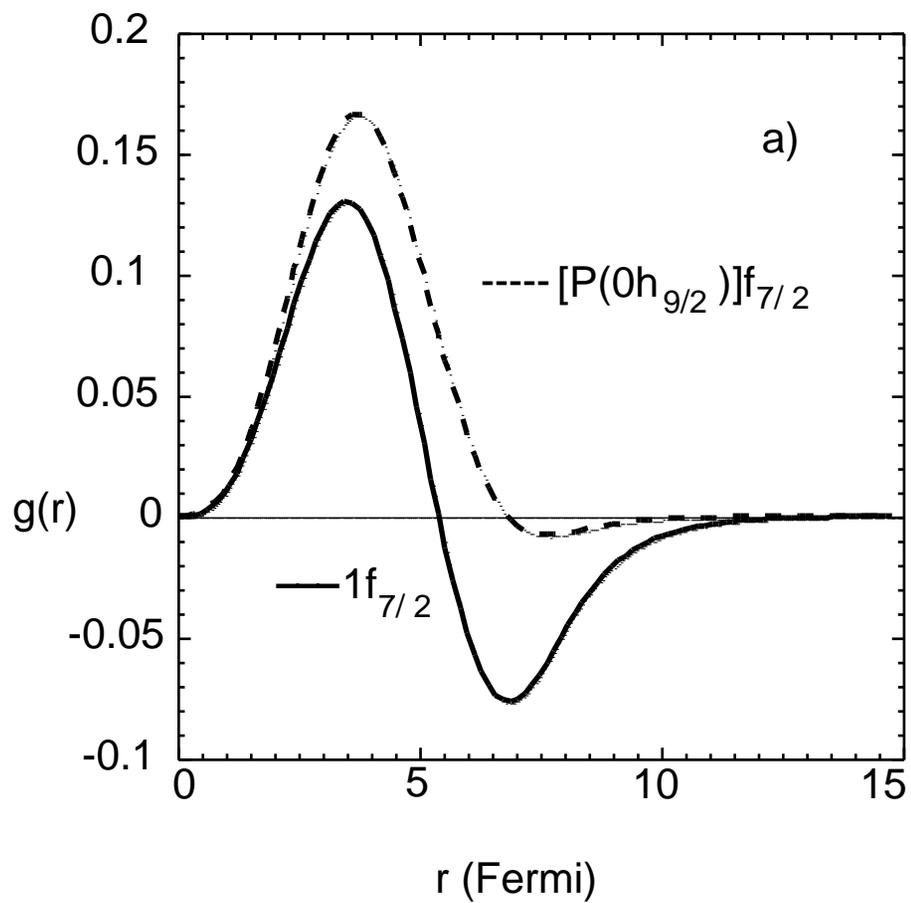
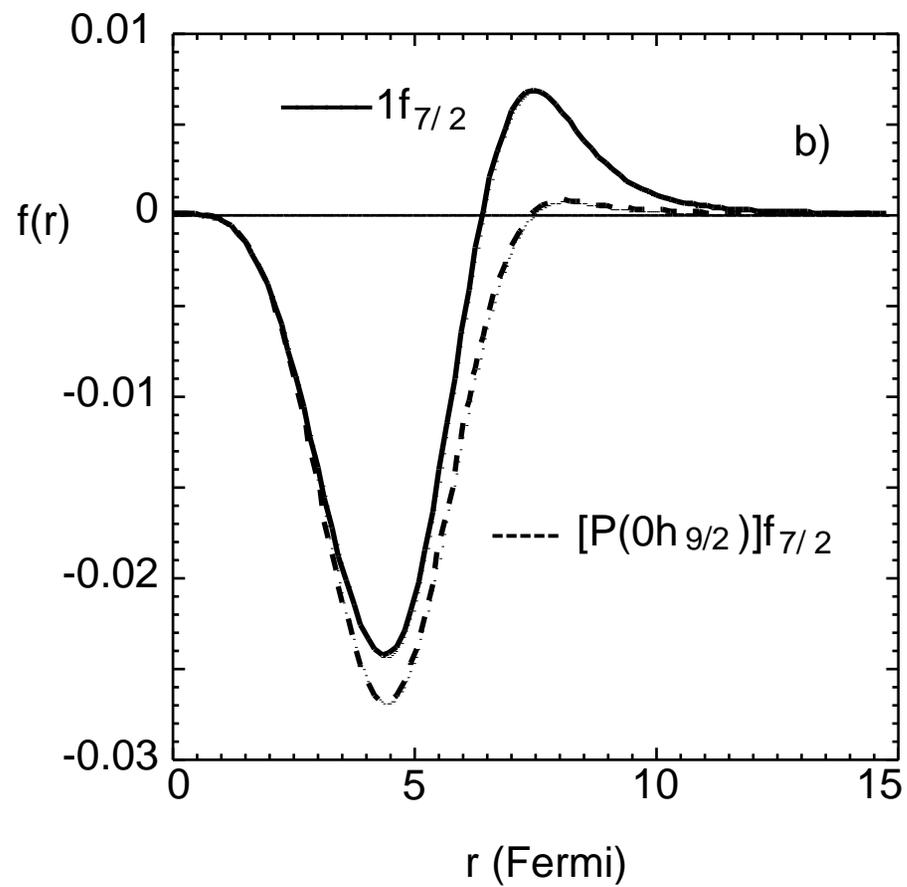
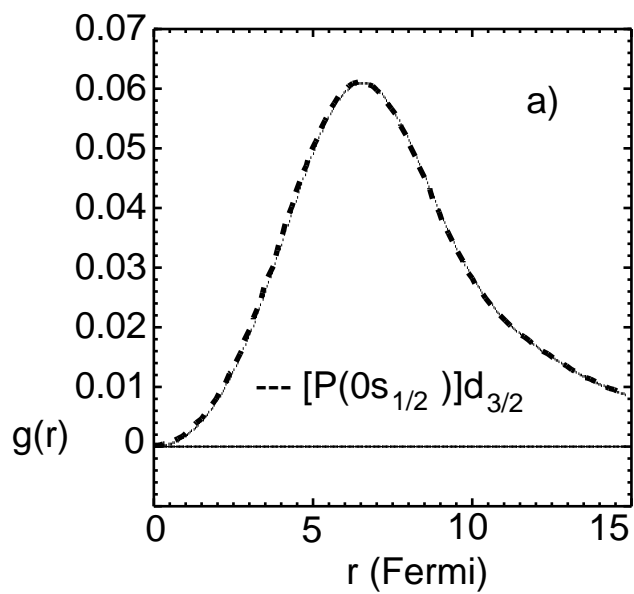
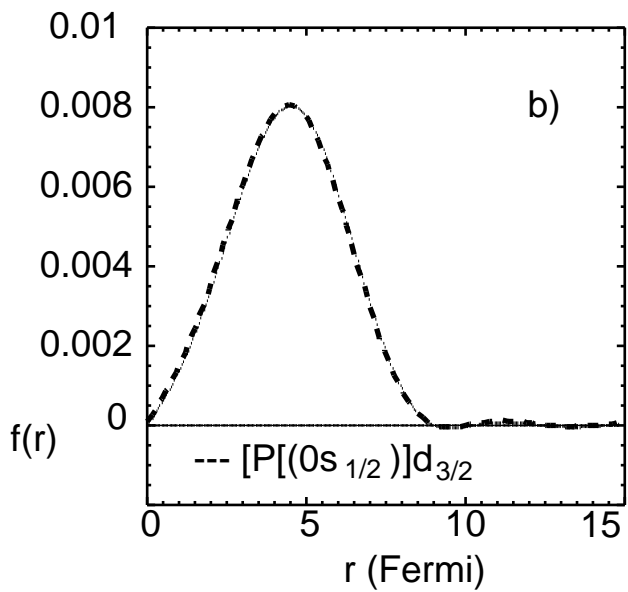
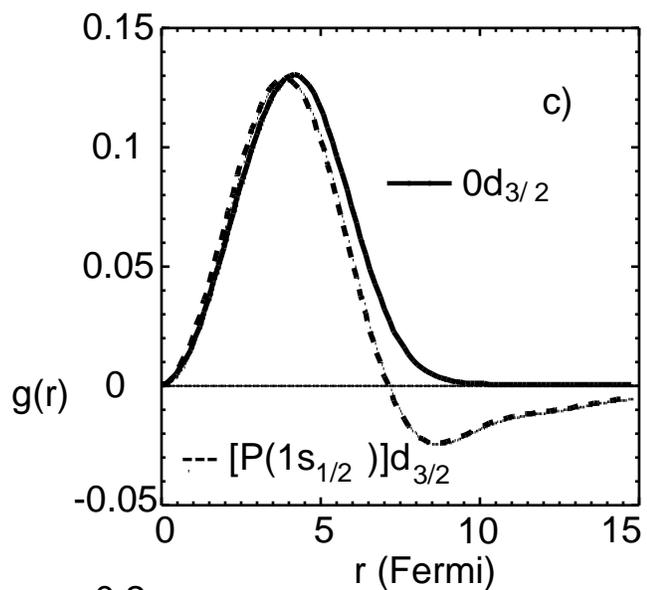
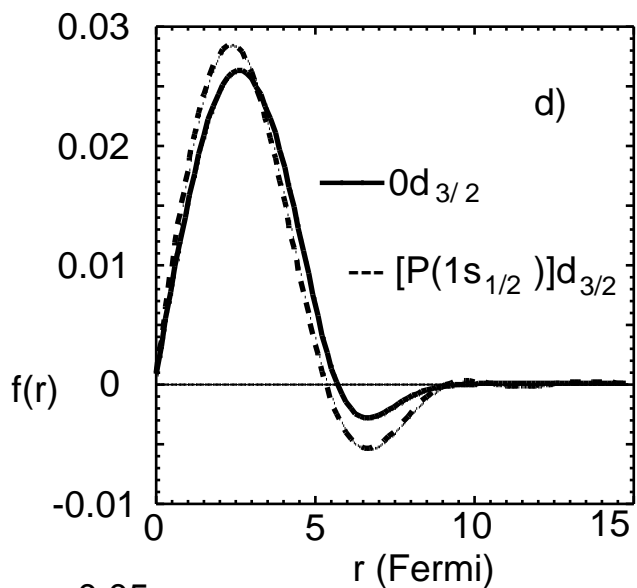
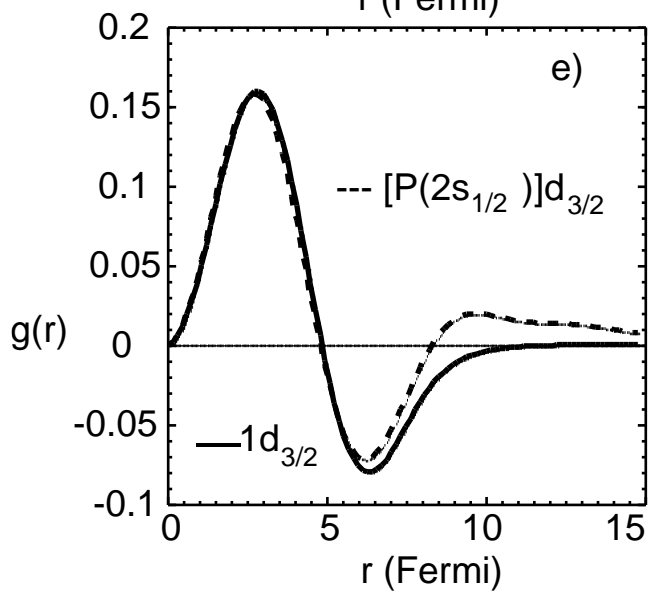
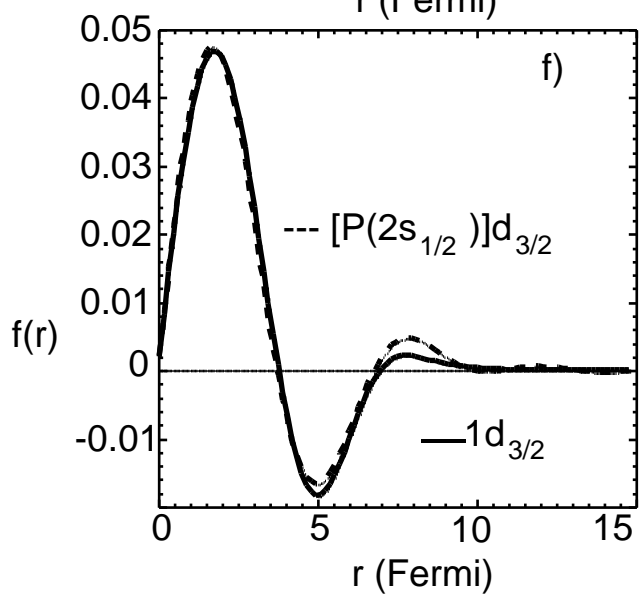
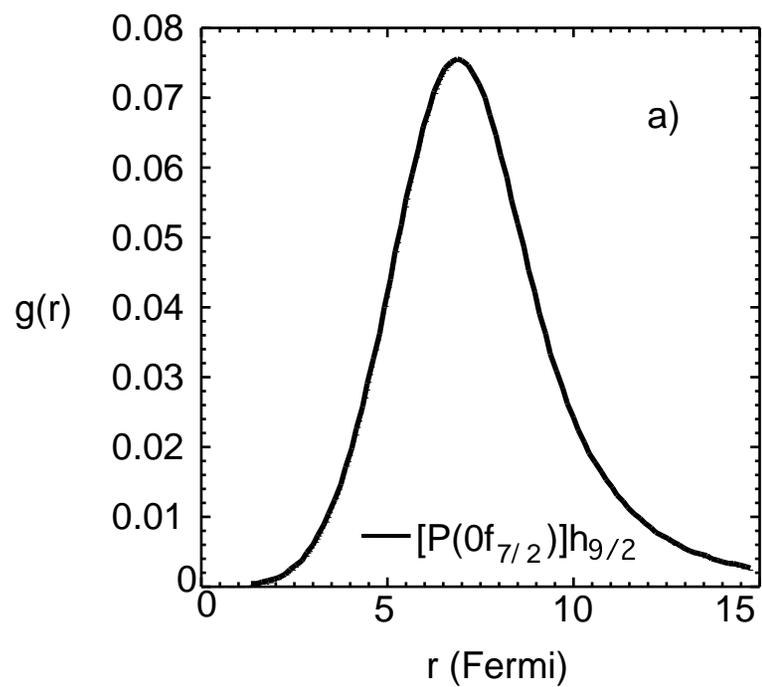
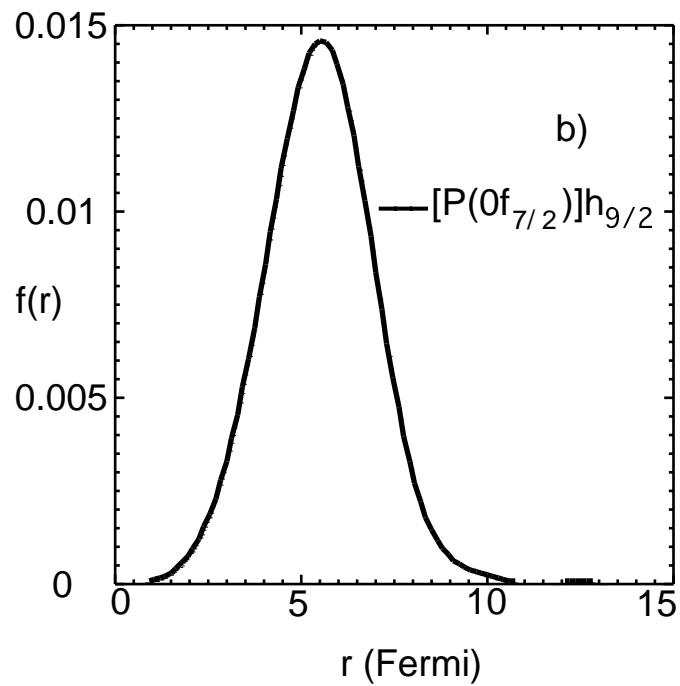
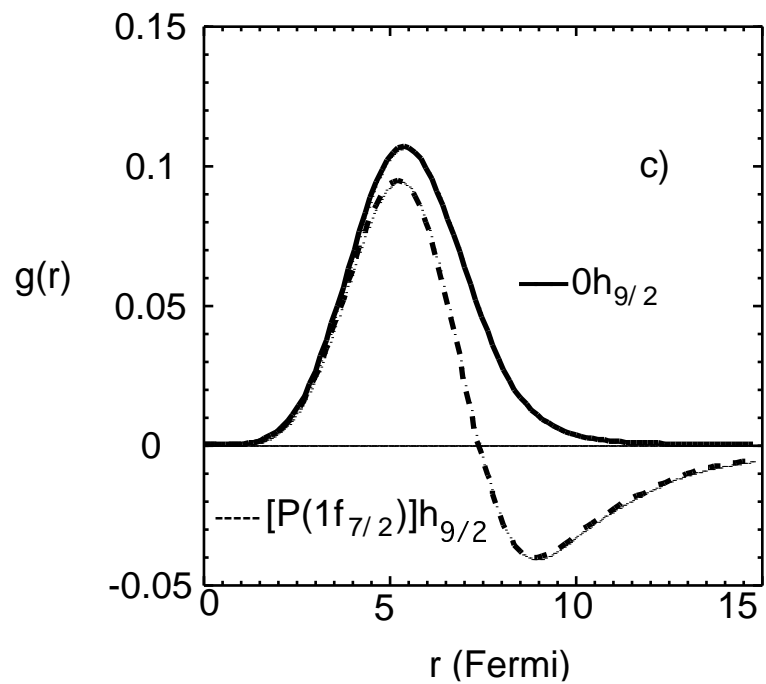
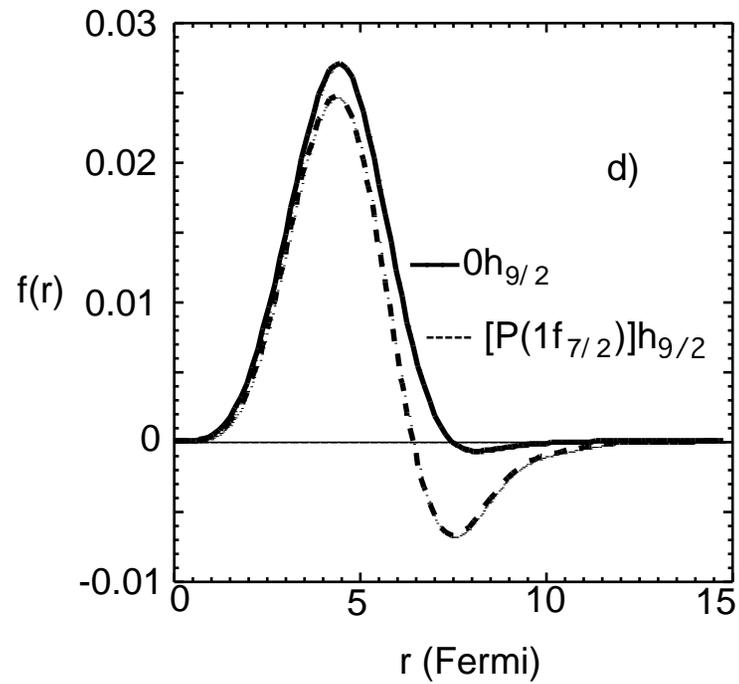
\end{document}